\documentclass[sigconf]{acmart}
\acmConference[EASE 2025]{The 29th International Conference on Evaluation and Assessment in Software Engineering}{17–20 June, 2025}{Istanbul, Türkiye}
\AtBeginDocument{%
  }

\setcopyright{acmlicensed}
\copyrightyear{2018}
\acmYear{2018}
\acmDOI{XXXXXXX.XXXXXXX}
\acmISBN{978-1-4503-XXXX-X/2018/06}

\usepackage{hyperref}
\usepackage{amsmath}
\usepackage{algorithm}
\usepackage{algorithmic}

\usepackage{url}

\usepackage{listings}

\definecolor{codegray}{gray}{0.95}
\definecolor{commentgreen}{rgb}{0,0.5,0}
\definecolor{keywordblue}{rgb}{0.13,0.13,1}
\definecolor{stringmauve}{rgb}{0.58,0,0.82}

\lstdefinelanguage{Solidity}{
  keywords={contract, function, public, view, returns, memory, address, string, uint256, require, emit, event, constructor, external, payable, return, bool, true, false},
  keywordstyle=\color{keywordblue}\bfseries,
  identifierstyle=\color{black},
  comment=[l]{//},
  commentstyle=\color{commentgreen},
  stringstyle=\color{stringmauve},
  sensitive=true,
  morecomment=[s]{/*}{*/},
}

\usepackage{listings}

\begin{document}

\title{Validation Framework for E-Contract and Smart Contract}





\author{Sangharatna Godboley, P. Radha Krishna, Sunkara Sri	Harika, and Pooja Varnam}
\affiliation{%
  \institution{NITMiner Technologies, Department of Computer Science and Engineering \\ National Institute of Technology Warangal,  \\}
  \city{Warangal, Telangana}
  \country{India}}
\email{{sanghu,prkrishna}@nitw.ac.in, {ss952022,vp842026}@student.nitw.ac.in.}

\renewcommand{\shortauthors}{Godboley et al.}

\begin{abstract}
We propose and  develop a framework for validating smart contracts derived from e-contracts. The goal is to ensure the generated smart contracts fulfil all the conditions outlined in their corresponding e-contracts. By confirming alignment between the smart contracts and their original agreements, this approach enhances trust and reliability in automated contract execution. The proposed framework will systematically compare and validate the terms and clauses of the e-contracts with the logic of the smart contracts. This validation confirms that the agreement is accurately translated into executable code. Automated verification identifies issues between the e-contracts and their smart contract counterparts. This proposed work will solve the problems of gap between legal language and code execution, this framework ensures seamless integration of smart contracts into the existing legal framework. 
\end{abstract}




\maketitle

\section{Introduction}
The rapid advancement of technology has led to a shift from traditional paper-based contracts to digital contracts, commonly known as e-contracts. These e-contracts provide the convenience of electronic storage, quick accessibility, and ease of sharing. However, with the rise of blockchain technology, a new type of contract has emerged Smart Contract. Smart contracts are self-executing agreements with the terms and conditions written directly in code. They offer automated enforcement, transparency and tamper-resistant execution. While smart contracts offer significant advantages over traditional contracts, their development process raises important concerns \cite{chen2021spesc,solrepSanghu,verisolmceSanghu}.

Smart contracts are created based on the terms and conditions specified in natural language within e-contracts \cite{upadhyay2020your}. The created smart contract code may be syntactically and logically correct, but it lacks the essential conditions outlined in the e-contract. The deployed smart contract may not meet the end user's expectations and requirements. This leads to potential contract failures and unsatisfactory outcomes that can lead to financial losses and legal disputes. So, one of the challenges is ensuring that smart contracts can be created and carried out under the party's willingness.

We aim to ensure that smart contracts function correctly and comply with legal regulations, addressing concerns about their security and reliability. Our research compares e-contracts and smart contracts using advanced techniques like pre-processing, dependency parsing, and constructing knowledge graphs. We analyse important entities in both contract types and identify any differences or discrepancies. 

For e-contracts, we refine the text and extract significant entities using pre-processing and dependency parsing. Next, we create a knowledge graph to get the relationships within the contract. Also, we generate knowledge graphs for smart contracts by generating Abstract Syntax Trees (AST) to extract information. By comparing the entities extracted from both types of contracts, we aim to highlight similarities, differences, and potential discrepancies. This research contributes to understanding how advanced techniques can improve the analysis of contracts, benefiting stakeholders in the legal and blockchain fields.

The rest of the paper is structured as follows: Section 2 deals with the Background, and Section 3 explains the Motivation. Section 4 shows the proposed approach with algorithms and frameworks. Section 5 discusses the experimental results. Finally, we conclude the paper and discuss future work in Section 6. 

\section{Background}
Natural Language Processing (NLP) techniques \cite{NLP1,NLP2} have been increasingly utilised in legal and contractual domains to extract meaningful insights from textual data. It is used to perform the automated analysis of e-contracts, extracting important entities, clauses, and relationships embedded within the text. The techniques, such as pre-processing, dependency parsing, and entity recognition, are used to examine the linguistic structure of e-contracts to get valuable information. 

Knowledge graph\cite{abdurahman2021lex2kg} generation plays a key role in enhancing the understanding of contractual agreements by representing entities and their relationships in a structured format. In e-contracts, knowledge graphs capture the semantics and connections between contractual clauses, commitments, and parties involved. Creating a knowledge graph based on the extracted entities and their dependencies, we get the relationships within e-contracts. This helps for a deeper understanding of the contractual responsibilities and commitment which facilitates the identification of discrepancies or inconsistencies between different clauses or parties. Knowledge graph generation is a powerful tool for visualising and navigating the complex landscape of contractual agreements, ultimately validating and verifying e-contracts against their corresponding smart contracts.

Literature shows innovative techniques such as the Software Word Usage Model (SWUM)\cite{ARTHUR20202522} in automatic comment generation for programming languages like Java. SWUM has shown promise in generating comments based on the usage of words within the code \cite{wohrer2018smart}. Applying a universal SWUM to languages like Solidity, specifically designed for blockchain operations, presents challenges. Solidity incorporates numerous reserved terms unique to blockchain, which may hinder the effectiveness of a generic SWUM. 

\section{Motivation}
We have proposed and designed a validation system for smart contracts and e-contracts. This will ensure the smart contracts' integrity, security, and functionality and will verify that they accurately represent the terms and conditions outlined in the e-contracts. 

We have identified some challenges in this domain.
\textbf{Complexity Discrepancy:} Smart contracts, implemented in languages like Solidity, can have complexities which have not been fully captured in their e-contract representations. This leads to discrepancies in functionality and execution.

\textbf{Regulatory Compliance:} It ensures that smart contracts adhere to legal regulations outlined in e-contracts, posing a significant challenge, particularly in jurisdictions with evolving blockchain and smart contract laws.

\textbf{Semantic Alignment:} Variations in language and structure between e-contracts and smart contracts may hinder accurate semantic alignment, making it difficult to verify if all contractual obligations are faithfully translated into executable code.
\section{Proposed Approach}
Our research employs advanced pre-processing, dependency parsing, and knowledge graph construction to analyze and compare e-contracts and smart contracts. Our approach involves extracting and analyzing important entities from both contract types.

\subsection{Algorithms}
\textcolor{black}{
	The proposed methodology begins with the construction of a knowledge graph from the natural language e-contract, as detailed in Algorithm~\ref{alg:e_contract_kg}. The process starts by preprocessing the e-contract text to eliminate noise such as special characters and irrelevant symbols, followed by tokenization and normalization. Using Natural Language Processing (NLP) techniques, particularly dependency parsing, key entities (e.g., parties, obligations, and dates) and their semantic relationships are extracted. These entities serve as the nodes and the relationships as the edges in the resulting knowledge graph $\mathcal{G}_e = (V_e, E_e)$. This graph-based structure enables a formal representation of the contract that is both interpretable and suitable for automated reasoning and visualization.}

\textcolor{black}{In parallel, a knowledge graph is also constructed from the smart contract, as described in Algorithm~\ref{alg:sc_kg}. The Solidity source code is analyzed by first extracting the compiler version and generating the Abstract Syntax Tree (AST) through compilation. The AST contains the hierarchical structure of the contract, including its functions, variables, and control logic. Using a custom grammar-driven approach, these elements are translated into human-readable descriptions that represent the functional semantics of the smart contract. Entities and dependencies derived from this semantic representation are used to build the smart contract knowledge graph $\mathcal{G}_s = (V_s, E_s)$. Finally, as shown in Algorithm~\ref{alg:compare_kgs}, the two knowledge graphs $\mathcal{G}_e$ and $\mathcal{G}_s$ are compared by matching their nodes and edges to identify similarities and discrepancies. This comparison as shown in Algorithm~\ref{alg:main}, provides a structured validation framework to ensure that the smart contract faithfully implements the specifications outlined in the original e-contract.}

	\begin{algorithm}[!]
		\caption{\textcolor{black}{E-Contract Knowledge Graph Construction}}
		\label{alg:e_contract_kg}
			\begin{algorithmic}[1]
				\REQUIRE Raw e-contract text $T_e$
				\ENSURE Knowledge graph $\mathcal{G}_e = (V_e, E_e)$
				
				\STATE $T_e' \gets \text{Preprocess}(T_e)$ \COMMENT{Remove noise, tokenize}
				\STATE $V_e \gets \text{ExtractEntities}(T_e')$
				\STATE $E_e \gets \text{ExtractRelations}(T_e')$
				\STATE $\mathcal{G}_e \gets (V_e, E_e)$
				\RETURN $\mathcal{G}_e$ 
			\end{algorithmic}
	\end{algorithm}
\vspace{-0.3cm}
	\begin{algorithm}[!]
		\caption{\textcolor{black}{Smart Contract Knowledge Graph Construction}}
		\label{alg:sc_kg}
			\begin{algorithmic}[1]
				\REQUIRE Solidity source code $S$
				\ENSURE Knowledge graph $\mathcal{G}_s = (V_s, E_s)$
				
				\STATE $v \gets \text{ExtractSolidityVersion}(S)$
				\STATE $\text{Compiler} \gets \text{SelectCompiler}(v)$
				\STATE $\text{AST} \gets \text{CompileAndGenerateAST}(S, \text{Compiler})$
				\STATE $\text{AST}_\text{json} \gets \text{SaveAsJSON}(\text{AST})$
				
				\FOR{each node $n \in \text{AST}$}
				\STATE $t \gets \text{NodeType}(n)$
				\STATE $d \gets \text{ApplyGrammar}(t)$
				\STATE Add description $d$ to semantic structure
				\ENDFOR
				
				\STATE $V_s \gets \text{ExtractEntitiesFromSemanticStructure}()$
				\STATE $E_s \gets \text{ExtractRelationsFromSemanticStructure}()$
				\STATE $\mathcal{G}_s \gets (V_s, E_s)$
				\RETURN $\mathcal{G}_s$
			\end{algorithmic}
	\end{algorithm}
\begin{algorithm}[!]
\caption{\textcolor{black}{Compare Knowledge Graphs}}
\label{alg:compare_kgs}
\begin{algorithmic}[1]
\REQUIRE $\mathcal{G}_e = (V_e, E_e)$, $\mathcal{G}_s = (V_s, E_s)$
\ENSURE Discrepancy report $\Delta$

\STATE $M_V \gets \text{MatchEntities}(V_e, V_s)$
\STATE $M_E \gets \text{MatchRelations}(E_e, E_s)$
\STATE $\Delta_V \gets V_e \cup V_s \setminus M_V$
\STATE $\Delta_E \gets E_e \cup E_s \setminus M_E$
\STATE $\Delta \gets (\Delta_V, \Delta_E)$
\RETURN $\Delta$
\end{algorithmic}
\end{algorithm}

\begin{algorithm}[!]
\caption{\textcolor{black}{Integrated Contract Analysis and Comparison}}
\label{alg:main}
\begin{algorithmic}[1]
\REQUIRE E-contract text $T_e$, smart contract code $S$
\ENSURE Validation report $\Delta$

\STATE $\mathcal{G}_e \gets \textsc{E-Contract Knowledge Graph Construction}(T_e)$ \COMMENT{Call Algorithm \ref{alg:e_contract_kg}}
\STATE $\mathcal{G}_s \gets \textsc{Smart Contract Knowledge Graph Construction}(S)$ \COMMENT{Call Algorithm \ref{alg:sc_kg}}
\STATE $\Delta \gets \textsc{Compare Knowledge Graphs}(\mathcal{G}_e, \mathcal{G}_s)$ \COMMENT{Call Algorithm \ref{alg:compare_kgs}}
\RETURN $\Delta$
\end{algorithmic}
\end{algorithm}

\subsection{Knowledge Graph construction of E-Contract}
\begin{figure}[h]
\centering
\includegraphics[height=0.31\textwidth,width=0.5\textwidth]{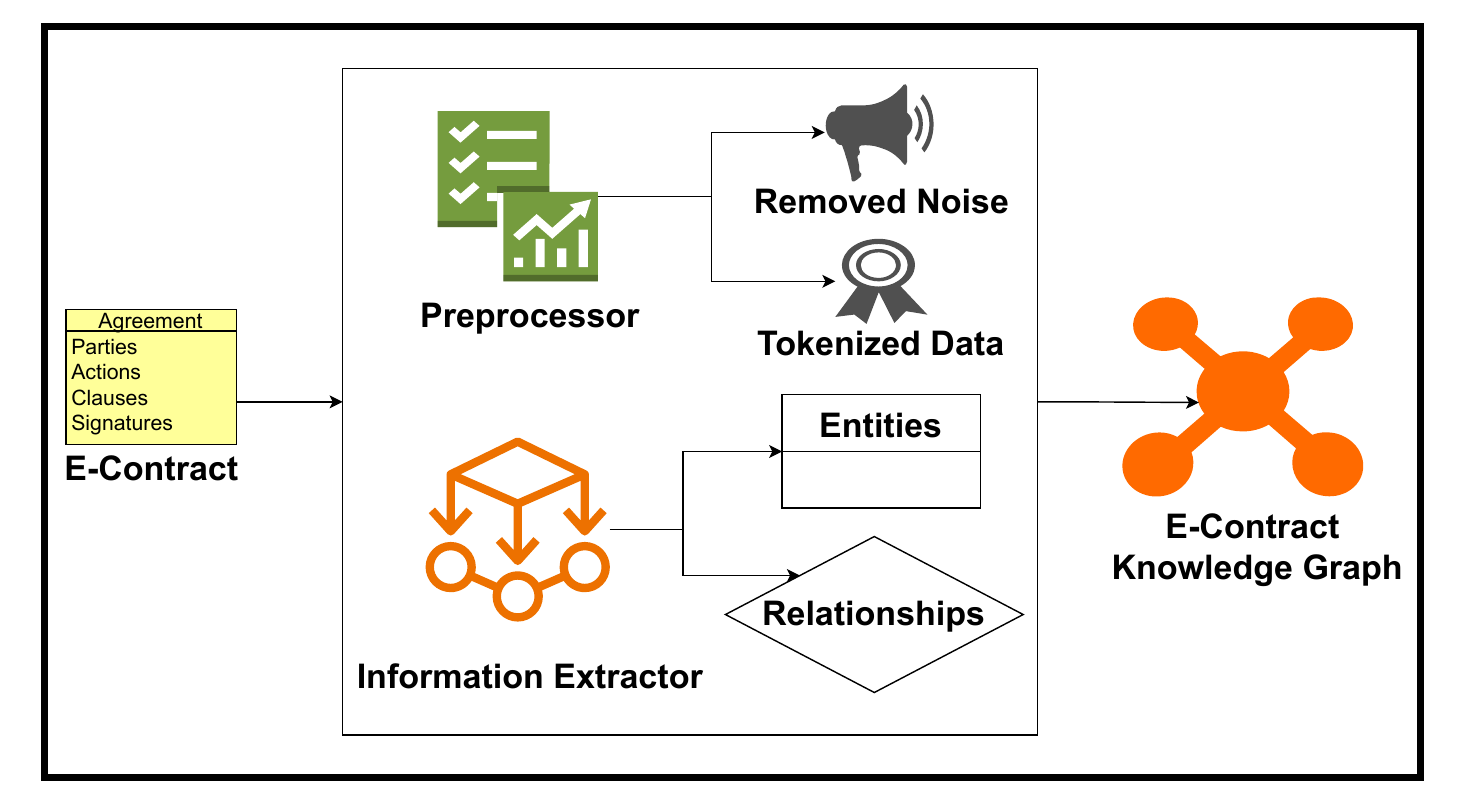}
\caption{Framework for generation of E-Contract Knowledge Graph}
\label{fig:eckgfm}
\end{figure}

Fig. \ref{fig:eckgfm} shows the framework for the generation of a knowledge graph from E-Contract.  Utilizing Natural Language Processing (NLP) techniques and dependency parsing, we generated a knowledge graph from e-contracts \cite{abdurahman2021lex2kg}. This process involved extracting entities and identifying relationships between them embedded within the contract's textual data. We used the NetworkX library\cite{hagberg2008exploring} to generate graphs and matplotlib\footnote{https://matplotlib.org/} to visualise the graphs. 
\begin{itemize}
    \setlength\itemsep{0em} 
    \item Preprocess the text to remove or modify any unwanted characters or symbols.
    \item Extract entities and identify the relation between them.
    \item Nodes contain entities, and the edges represent the relation between the nodes.
\end{itemize}

\subsection{Knowledge Graph Generation for Smart Contract}
\begin{figure}[h]
\centering
\includegraphics[height=0.34\textwidth,width=0.5\textwidth]{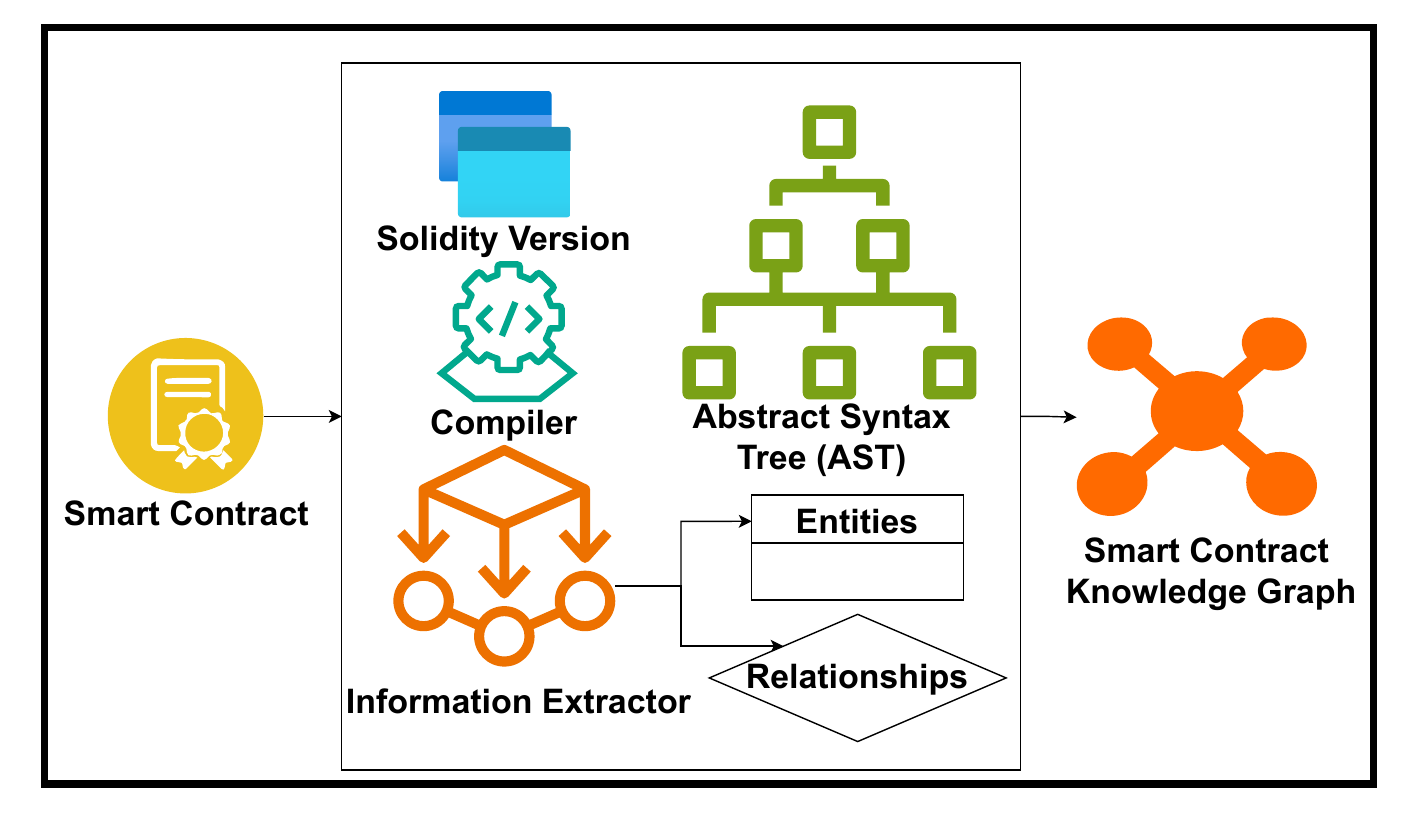}
\caption{Framework for generation of  Smart Contract Knowledge Graph}
\label{fig:sckgfm}
\end{figure}

Fig. \ref{fig:sckgfm} shows the Knowledge Graph Generation for Smart Contract into the Knowledge Graph. 

Abstract Syntax Tree (AST) generation begins by reading the contents of the specified Solidity smart contract file \cite{shia2021solctrans}. It then extracts the Solidity version indicated in the contract's pragma statement, which is crucial for selecting the compatible Solidity compiler version. The extracted solidity version is utilized, and the program selects the appropriate Solidity compiler version to compile the smart contract and avoid compile time issues. 

Following the compilation process, the program captures the compiler output, including any developer documentation generated during compilation. This documentation contains detailed insights into the contract's methods and functions, including their descriptions and parameters \cite{chen2021spesc}.

After parsing the developer documentation from the compiler output, the program generates the Solidity smart contract's Abstract Syntax Tree (AST). The AST represents the smart contract's structure and syntactic elements. It is saved in JSON format, which helps understand the smart contract's structure and supports code review and debugging. The program automatically creates the AST, which makes analyzing and validating smart contracts faster and more accurate.

\par In our approach, we parsed the generated Abstract Syntax Tree (AST) based on node types, utilizing this structured representation of the contract's syntax to facilitate the translation of code into human-readable language \cite{shia2021solctrans}. We have designed a grammar that is connected to the specific node types encountered in the AST. This decodes the smart contract's logic and functionality systematically. We developed a framework for interpreting the contract's syntax by classifying the nodes based on their types, such as variables, functions, conditionals, and loops.

For instance, variables and their attributes were mapped to grammar rules specifying their data types, names, and initialization values. The functions were translated into human-readable descriptions, which described their parameters, return types, and operations within the function body. The conditional statements and loops were parsed to capture their logical conditions and iterative behaviours. This shows clear explanations of control flow within the smart contract. We were able to systematically traverse the AST by connecting to this grammar-driven approach. We extracted relevant information from each node and composed readable descriptions of the smart contract's structure and behaviour. The stakeholders understand its logic and implications in a human-readable format. Additionally, it enhanced the accessibility of smart contract code, enabling non-technical stakeholders to engage meaningfully with the contract's contents and implications.

\begin{lstlisting}[caption={E-Contract for Rental Agreement}, label={lst:rental_E-contract}]
Parties: This Smart Rental Contract (the "Contract") is entered into on 01/01/2025, between.

Landlord: ABC, residing in London, United Kingdom 

Tenant: XYZ, residing at Singapore, Singapore 

Property: The Landlord agrees to rent the following property to the Tenant:

Address: 123 Baker Street, London, NW1 4NW 

Term: The term of this Contract shall be from 01/01/2025 to 31/12/2025.

Rent: The Tenant agrees to pay a monthly rent of GBP 5000. Rent is due on the first day of each month.

Security Deposit: The tenant shall deposit GBP 2000 as a security deposit upon signing this Contract. The security deposit will be held by the Landlord and returned to the Tenant upon termination of the Contract, subject to any deductions for damages or unpaid rent.

Utilities: The Tenant is responsible for paying all utility charges, including electricity, water, gas, and internet.

Use of Property: The Tenant shall use the property solely as a private residence and shall not engage in any unlawful activities on the premises.

Maintenance and Repairs: The Landlord is responsible for maintaining the property in good condition. The Tenant shall keep the property clean and sanitary.

Termination: Either party may terminate this Contract with one month written notice to the other party.

Governing Law: This Contract shall be governed by the laws of London, UK.

Signatures: In witness whereof, the parties hereto have executed this Contract on the date first above written.

[Landlord's Signature]: Date: 01/01/2025 ABC

[Tenant's Signature]: Date: 01/01/2025 XYZ

\end{lstlisting}

Our methodology used the parsed data, representing the smart contract in natural language, to construct a new knowledge graph. This knowledge graph encapsulates the semantic relationships and dependencies extracted from the human-readable representation of the contract's logic and functionality.

Once both knowledge graphs—derived from the original e-contract and the parsed smart contract—are constructed, we compare them systematically based on identifying similarities, differences, and potential discrepancies between the entities, relationships, and dependencies in the knowledge graphs.

\begin{lstlisting}[caption={Solidity Smart Contract for Rental Agreement}, label={lst:rental_contract}]
pragma solidity ^0.8.0;
contract RentalAgreement {
    address public landlord;
    address public tenant;
    string public propertyAddress;
    uint256 public rentAmount;
    uint256 public securityDeposit;
    uint256 public termStartDate;
    uint256 public termEndDate;
    bool public terminated;
    event RentPaid(uint256 amount);
    event SecurityDepositPaid(uint256 amount);
    event ContractTerminated(address terminatedBy);
    constructor(
        address _landlord,
        address _tenant,
        string memory _propertyAddress,
        uint256 _rentAmount,
        uint256 _securityDeposit,
        uint256 _termStartDate,
        uint256 _termEndDate
    ) {
        landlord = _landlord;
        tenant = _tenant;
        propertyAddress = _propertyAddress;
        rentAmount = _rentAmount;
        securityDeposit = _securityDeposit;
        termStartDate = _termStartDate;
        termEndDate = _termEndDate;
        terminated = false;
    }

    function payRent() external payable {
        require(msg.sender == tenant, "Only tenant can pay rent");
        require(msg.value == rentAmount, "Incorrect rent amount");
        emit RentPaid(msg.value);
    }

    function paySecurityDeposit() external payable {
        require(msg.sender == tenant, "Only tenant can pay security deposit");
        require(msg.value == securityDeposit, "Incorrect security deposit amount");
        emit SecurityDepositPaid(msg.value);
    }

    function terminateContract() external {
        require(msg.sender == landlord || msg.sender == tenant, "Unauthorized");
        terminated = true;
        emit ContractTerminated(msg.sender);
    }

    function getContractDetails() external view returns (
        address,
        address,
        string memory,
        uint256,
        uint256,
        uint256,
        uint256,
        bool
    ) {
        return (landlord, tenant, propertyAddress, rentAmount, securityDeposit, termStartDate, termEndDate, terminated);
    }
}
\end{lstlisting}

This proposed validation approach certifies that the smart contract accurately shows the intentions outlined in its e-contract. By using knowledge graphs, we enhance the reliability of automated smart contract execution systems that builds trust among stakeholders in smart contract integrity.

\section{Experimental Results}
\textcolor{black}{In this section, we discuss with a working example and show the results generated by our proposed approach. }
\subsection{Working Example}
\textcolor{black}{An e-contract, or electronic contract, is a digital version of a traditional agreement between two or more parties. In the rental agreement example, as shown in Listing \ref{lst:rental_E-contract}, the landlord and tenant agree to specific terms such as the rental period, monthly rent, and responsibilities like paying utilities and maintaining the property. In place of signing a traditional contract on paper approved by court of law, both parties can review and approve the contract electronically, making the process faster and more convenient. }

\textcolor{black}{A smart contract takes this step further by using computer code to enforce the agreement automatically. The rental agreement is written using a programming language called Solidity and deployed on a blockchain, as shown in Listing \ref{lst:rental_contract}. In this version, the contract stores all key details, such as the landlord's and tenant's addresses, the rent amount, the security deposit, and the rental period. Once the smart contract is created, it acts like a digital agent that can receive rent payments, hold deposits, and track whether the contract has been ended.}

\textcolor{black}{The smart contract also includes built-in rules that ensure only the tenant can make payments and that the correct amount is sent. It keeps track of payments and emits events—like a receipt—when rent or a deposit is paid. Importantly, it also allows either party to end the contract by calling a function. All actions taken on the contract are recorded on the blockchain, which means they are secure and transparent and cannot be changed later without everyone's knowledge.}

\textcolor{black}{This setup improves trust between the landlord and tenant since the rules are programmed and cannot be altered unfairly. There's no need for a middleman like a rental agency to manage the process. Smart contracts also reduce the chance of disputes, as everything is recorded clearly and handled automatically. As more people use blockchain in everyday transactions, smart contracts are becoming a reliable and efficient way to manage digital agreements.}


\begin{figure}[ht]
\centering
\boxed{\includegraphics[width=0.44\textwidth]{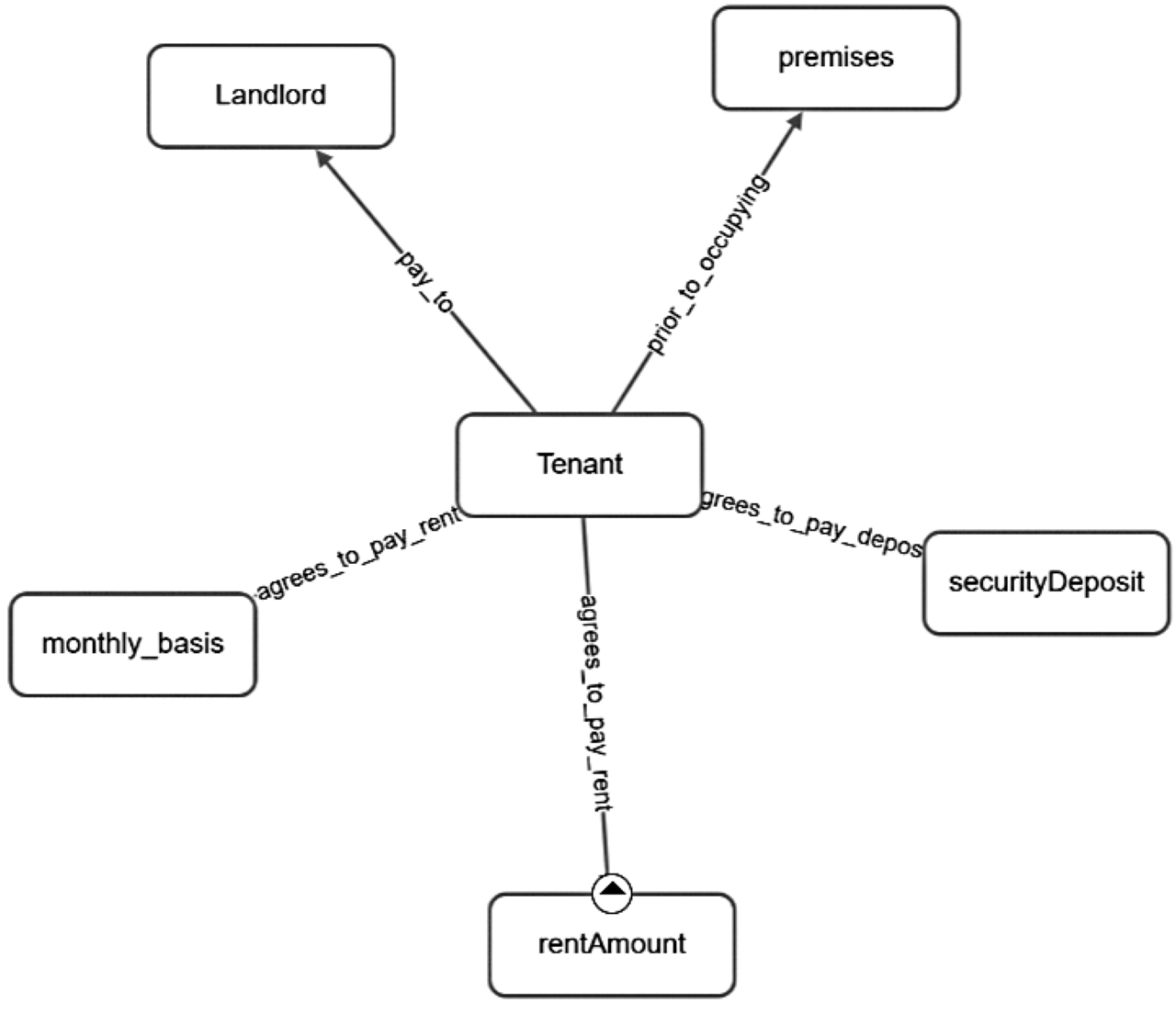}}
\caption{Knowledge graph of e-contract}
\label{fig:eckg}
\end{figure}


\begin{figure}[!]
\centering
\boxed{\includegraphics[width=0.44\textwidth]{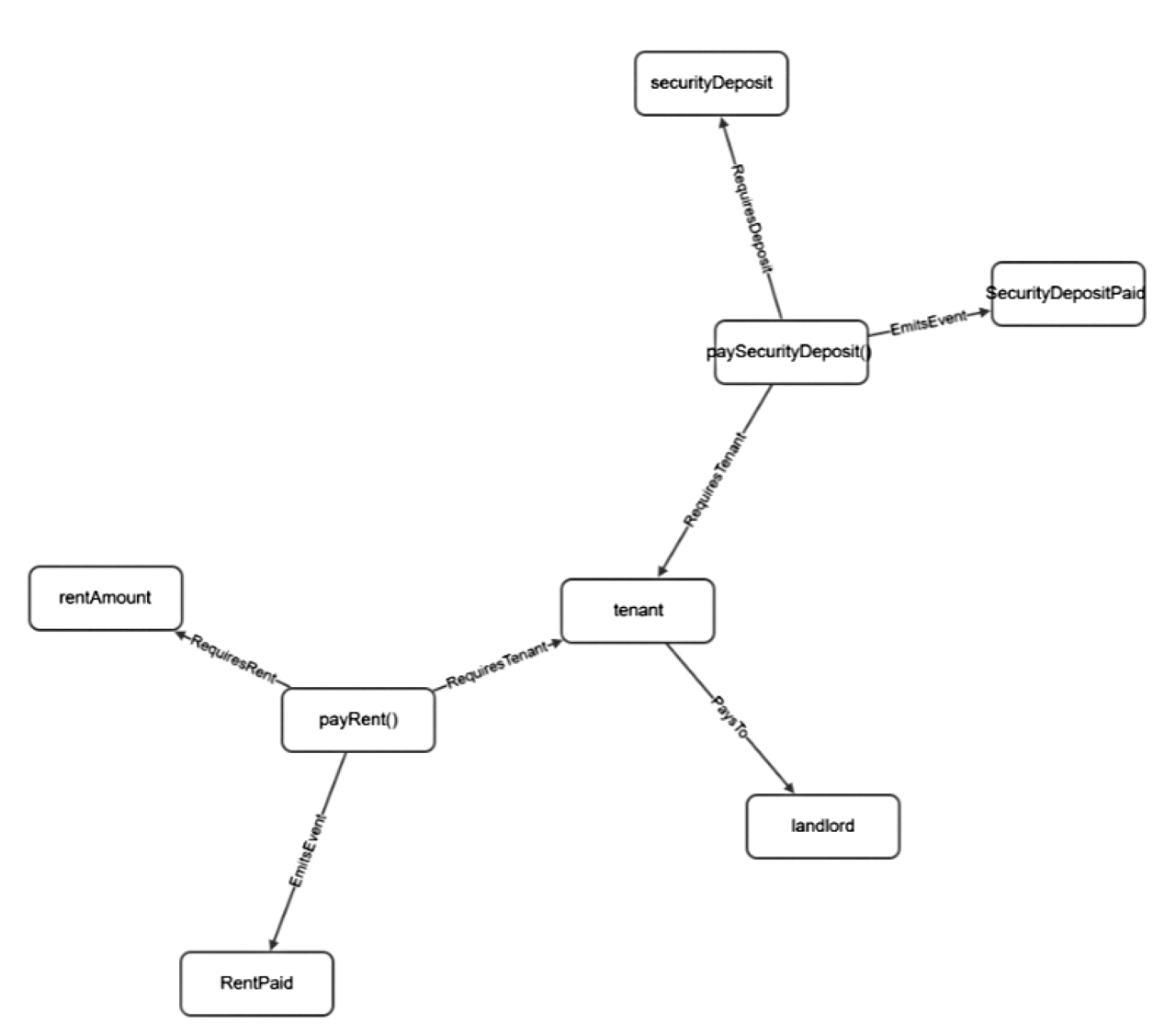}}
\caption{Knowledge graph of Smart Contract}
\label{fig:sckg}
\end{figure}

\begin{figure*}[!h]
\centering
\includegraphics[width=0.75\textwidth]{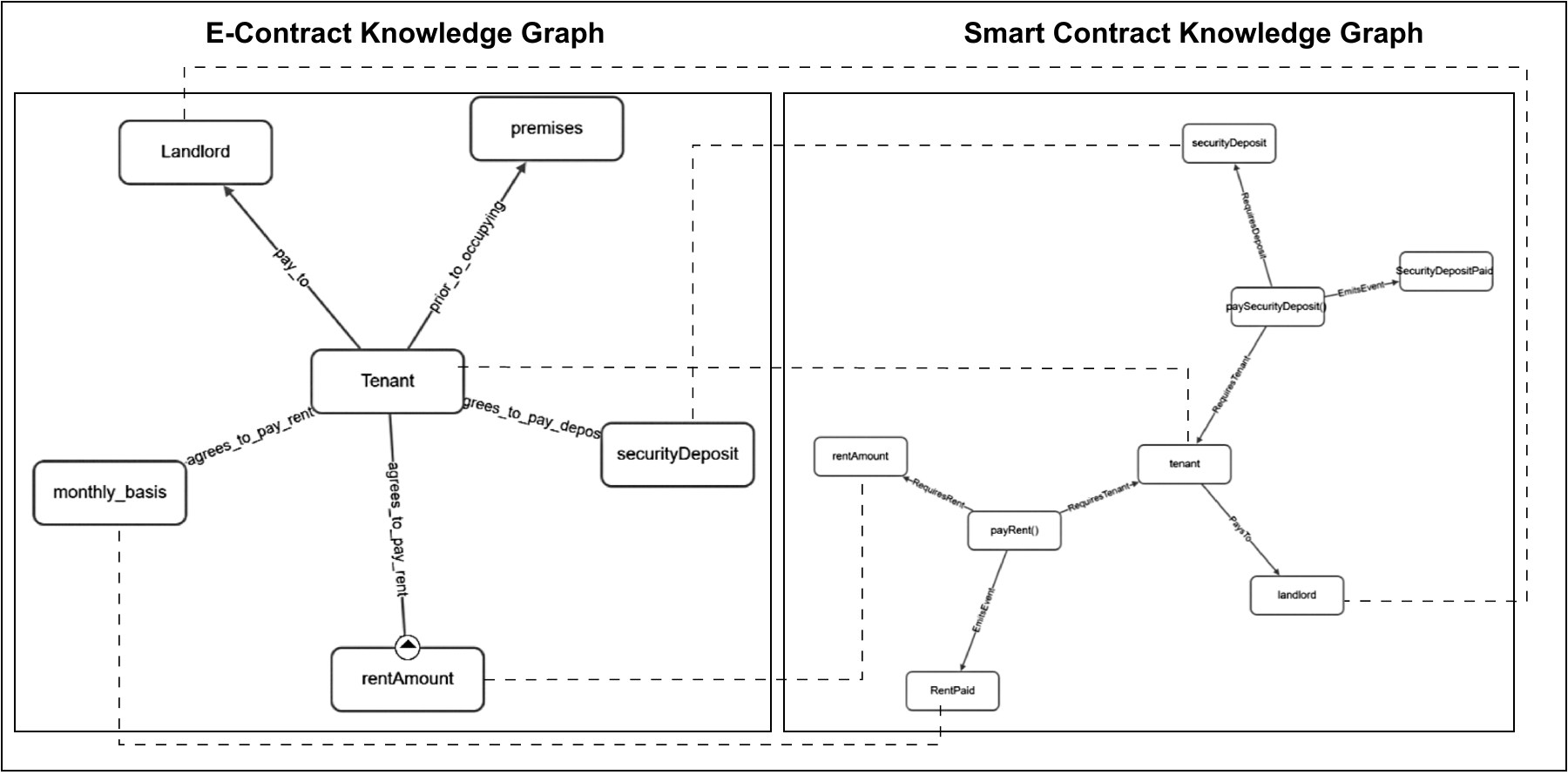}
\caption{Validation of generated knowledge graphs}
\label{fig:compecsc}
\end{figure*}

\begin{figure}[h]
\centering
\includegraphics[width=0.47\textwidth]{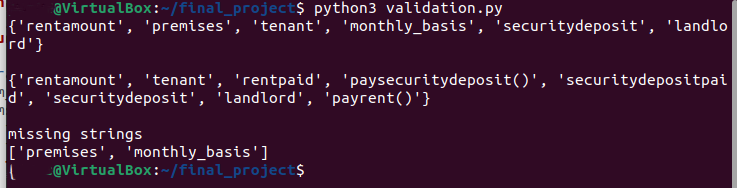}
\caption{Results after the comparison of json files}
\label{fig:validation}
\end{figure}

\textcolor{black}{Using the framework for generation of  E-Contract Knowledge Graph shown in Fig. \ref{fig:eckgfm} on E-Contract shown in Listing \ref{lst:rental_E-contract}, we generate the knowledge graph as shown in Fig. \ref{fig:eckg}. Similarly, using the framework for generation of Smart Contract Knowledge Graph shown in Fig. \ref{fig:sckgfm} on Smart Contract shown in Listing \ref{lst:rental_contract}, we generate the knowledge graph as shown in Fig. \ref{fig:sckg}.}

\subsection{Results}
For the validation phase, we process JSON files to compare data extracted from knowledge graphs representing the e-contract and the smart contract. Initially, we extracted pertinent information from each knowledge graph, encapsulating entities, relationships, and other relevant details. We then compared the JSON files generated from the two knowledge graphs. This comparative analysis aimed to discern discrepancies, differences, or inconsistencies between the contractual obligations outlined in the e-contract and their representation in the smart contract \cite{magazzeni2017validation}.

\textcolor{black}{Through this comparison process, we meticulously scrutinised whether the JSON files were identical, highlighting any variations between them. Subsequently, we meticulously examined the results to pinpoint any missing terms or entities in the e-contract or the smart contract knowledge graph, as shown in Fig. \ref{fig:validation}. We compare both the knowledge graphs generated for E-Contract and Smart Contract as shown in Fig. \ref{fig:compecsc}, and observe that most of the entities present in E-Contract are available in Smart Contract with the relationships. We have shown the mapping with a dashed line (----) connecting the same entities. }

\section{Conclusion and Future Work}
In this paper, we addressed the challenge of aligning e-contracts written in natural language with smart contracts written in solidity language. We focused on analyzing and comparing these contract types using advanced preprocessing, dependency parsing, and knowledge graph construction techniques. By extracting entities and building knowledge graphs, we aimed to identify discrepancies or differences in representation between e-contracts and smart contracts.

\textcolor{black}{In the future, this work could be expanded by converting e-contracts into Abstract Syntax Trees (ASTs) and comparing them with the ASTs of their related smart contracts. This would involve creating programs that can read and break down e-contracts into structured forms, similar to how smart contracts are represented. After generating both the versions in AST format, they can be validated to see if the smart contract accurately reflects the terms written in the e-contract. Adding this kind of comparison would make the whole process of validating digital contracts more thorough and trustworthy.}




\section*{Acknowledgement}
This work is sponsored under the grant DIAL Scheme, DST, Government of India given to NITMINER Tech. Pvt. Ltd. All rights reserved between NITMINER Tech. Pvt. Ltd. and IBITF, IIT Bhilai, Government of India.

\bibliographystyle{plain}
\bibliography{sample-base}

\end{document}